# Probing the p-Ge$_{1-x}$Si$_x$/Ge/p-Ge$_{1-x}$Si$_x$ quantum well by means of the quantum Hall effect

Yu G Arapov, G I Harus, V N Neverov, N G Shelushinina, M V Yakunin*,
G A Alshanskii, O A Kuznetsov$^\#$

Institute of Metal Physics RAS, Ekaterinburg, GSP-170, 620219, Russia
$^\#$Scientific-Research Institute at Nizhnii Novgorod State University, Russia

**Abstract.** We have measured the temperature ($0.1 \leq T \leq 15$ K) and magnetic field ($0 \leq B \leq 32$ T) dependences of longitudinal and Hall resistivities for the p-Ge$_{0.93}$Si$_{0.07}$/Ge multilayers with different Ge layer widths $10 \leq d_w \leq 38$ nm and hole densities $p_s = (1 \div 5) \cdot 10^{15}$ m$^{-2}$. An extremely high sensitivity of the experimental data (the structure of magnetoresistance traces, relative values of the inter-Landau-level (LL) gaps deduced from the activated magnetotransport etc) to the quantum well (QW) characteristics has been revealed in the cases when the Fermi level reached the second confinement subband. The background density of states $(5 \div 10) \cdot 10^{14}$ m$^{-2}$meV$^{-1}$ deduced from the activation behavior of the magnetoresistance was too high to be attributed to the LL tails, but may be accounted for within a smooth random potential model. The hole gas in the Ge quantum well was found to separate into two sublayers for $d_w > \sim 35$ nm and $p_s \approx 5 \cdot 10^{15}$ m$^{-2}$. Concomitantly the positive magnetoresistance emerged in the weakest fields, from which different mobilities in the sublayers were deduced. A model is suggested to explain the existence of the plateaux close to the fundamental values in a system of two parallel layers with different mobilities.

## 1. Introduction

The p-type modulation-doped heterostructures on the basis of Si, Ge and their alloys are of interest in device physics, in view of Si-based chip technology for fabrication of high performance transistors, intra-chip optical interconnects and possible applications for optical-fiber telecommunications. On the other hand, many quantum physical aspects possibly will be realized in investigations on this heterosystem due to specific features of the valence band structure, which might be varied dramatically by the extent of hole confinement and the uniaxial stress arising from the lattice mismatch. Since the band offset in the Ge-Si heterosystem is almost entirely located in the valence band, the confinement of electrons is hardly achievable and 2D conductance is mainly due to holes.

The majority of research has been performed so far on the Si-side compounds, like Si/Si$_{1-x}$Ge$_x$/Si quantum well (QW), to be best compatible with the Si-based technology. In this case the hole conductivity is via the Si-Ge alloy. A two-dimensional hole gas (2DHG) of better quality, with low temperature mobilities up to 1.7 m$^2$/V·s, has been achieved in the heterosystems on the Ge side since in this case the 2DHG is confined not in an alloy with randomly distributed Ge and Si atoms within crystal lattice but in a uniform Ge crystal.

In this paper we report on the investigations of the quantum magnetotransport in multilayered modulation doped p-Ge$_{1-x}$Si$_x$/Ge heterostructures with a small amount of Si. We know of just a limited number of experiments on quantum magnetotransport of holes in 2D Ge [1,2]. An essential feature of these works was that the potential profile of a Ge QW was skewed either due to single-side doping [1] or to outside electric field applied through a front gate [2]. The nominally symmetric potential profile associated with the Ge layers in our samples has made it possible to study the effects of the hole distribution in the QW cross section as a function of QW width and 2DHG density.

Some details of the sample preparation will be described in section 2. In section 3 the results will be given for the exploration of p-Ge$_{1-x}$Si$_x$/Ge/p-Ge$_{1-x}$Si$_x$ QWs within the range of widths and hole densities corresponding to the integrated hole gas in the Ge layer and peculiarities due to involvement of the second confinement subband will be analyzed. In addition, the advantages of the activation analysis for the description of Landau level (LL) picture will be given and background density of states (DOS) estimated. In section 4 the phenomenon of the hole gas separation into two 2D sublayers inside sufficiently wide and populated Ge layers will be described as well as the consequences that reflect this process in the quantum magnetotransport.

* To whom correspondence should be addressed.
E-mail: yakunin@imp.uran.ru

## 2. Experimental technique

A series of multilayered p-$Ge_{1-x}Si_x$/Ge (x ≈ 0.07) heterostructures differing in the Ge layer width in the range $d_w$ = 12 ÷ 40 nm and the hole density per single Ge layer $p_s$ = (1 ÷ 5)·$10^{15}$ $m^{-2}$ were grown by hydride vapor deposition on the Ge(111) substrate. The undoped Ge buffer was grown first, followed by the undoped $Ge_{1-y}Si_y$ buffer and several undoped $Ge_{1-x}Si_x$/Ge periods. The relation between Si content in the buffer (y) and in the multilayers (x) presets the distribution of mismatch stress between the Ge and $Ge_{1-x}Si_x$ layers. The current flowed via the part of the multilayered structure where the $Ge_{1-x}Si_x$ barriers were selectively doped with boron in their central parts (with undoped spacers about ¼ of the barrier width left on both sides of the barriers): see the inset in figure 2. The barriers were sufficiently wide to avoid the inter-Ge-layer tunneling. The low temperature hole mobilities were in the range 1 ÷ 1.7 $m^2$/V·s.

The double cross Hall bridges were fabricated by the photolithography and subsequent wet etching technique and contacts attached by the thermocompression.

Hall and longitudinal magnetoresistivities were measured on dc currents in normal magnetic fields up to 12 T in the steady regime and up to 32 T in ~10 ms pulses within the temperature range 0.1 ≤ T ≤ 15 K.

**Table 1.** Sample parameters.

| Sample | $\mu$, $m^2$/V·s | $p_s$, $10^{15}$ $m^{-2}$ | $d_w$,* nm | $p_s d_w^2$ |
|---|---|---|---|---|
| 1006-1 | 1.4 | 4.9 | 12.5 | 0.77 |
| 1124b3 | 1.0 | 2.8 | 20(21.4) | 1.28 |
| 1125a7 | 1.7 | 2.8 | 20(22) | 1.36 |
| 1123a6 | 1.4 | 3.4 | 20(23.5) | 1.88 |
| 1003-2 | 1.5 | 4.8 | 22 | 2.32 |
| 451b4 | 1.4 | 1.4 | 35.5 | |
| 476a4 | 0.8 | 5 | 38 | |

*In brackets we present the corrected values obtained from our analysis.

## 3. Integrated hole gas in a Ge layer

### 3.1. Manifestations of the second subband in $\rho_{xx}(B)$ and $\rho_{xy}(B)$ experimental traces

The simplest situation occurs for the narrowest Ge layers with the lowest hole densities. In this case the QW profile approaches a rectangular shape and the lowest confinement subband only is filled with holes. Though a system of Landau levels in the valence band is rather complicated [3] due to heavy and light hole hybridization complemented by the effects of confinement and uniaxial stress (fig.1b), the experimental recordings of the longitudinal and Hall magnetoresistivities, $\rho_{xx}(B)$ and $\rho_{xy}(B)$, have a regular structure similar to that observed in a simple undegenerate conduction band (fig.1a). The $\rho_{xy}(B)$ traces contain plateaus of integer quantum Hall effect (QHE) at fundamental values $\rho_{xy} = h/ie^2$ (where i is an integer value) concomitant with minima in $\rho_{xx}(B)$, both kinds of peculiarities being regularly spaced in reciprocal magnetic field and monotonously damped with decreasing field. The peculiarities correspond to i = 1, 2, 4, 6…, i.e. the even numbered QH peculiarities dominate at weak fields. This is in analogy with a simple conductivity band for small Zeeman splitting.

Magnetoresistivities of quite different structure were detected for the samples with wider Ge layers and higher hole densities [4]. In figures 2 and 3 the Hall and longitudinal magnetoresistivities are presented for the samples in the first five rows of the table. For longitudinal magnetoresistivity the traces are normalized in their amplitudes to the highest peak and in their positions by scaling them versus inverse filling factor $\nu = p_s h/eB$. The general feature of these five samples is that, in spite of different QW widths and hole densities, their traces contain plateaux in $\rho_{xy}(B)$ and concomitant $\rho_{xx}(B)$ minima for i = 1 and 2, but the structure of these curves differs significantly below the plateau with i = 2.

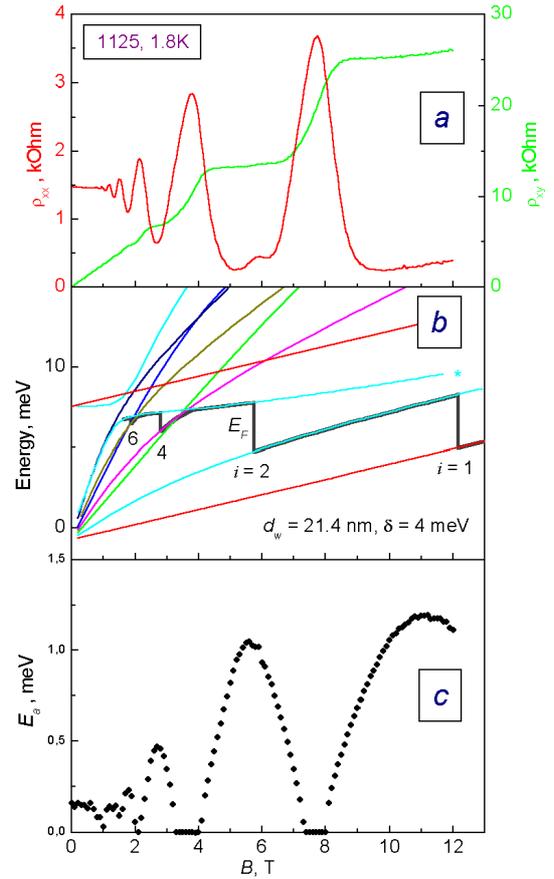

**Figure 1.** Comparison of the calculated LL picture and the Fermi level motion (b) with the experimental recordings (a) and deduced activation energies (c) for the sample 1125a7. Mobility gaps are estimated from (c) as twice the maximum $E_a(B)$ values.

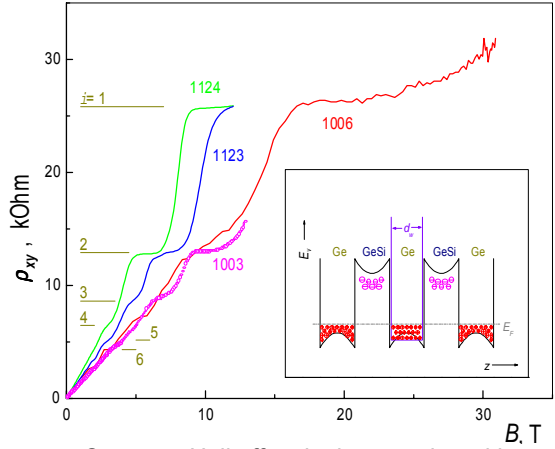

**Figure 2.** Quantum Hall effect in the samples with integrated hole gas. Inset: a schematic valence band energy diagram of the samples studied.

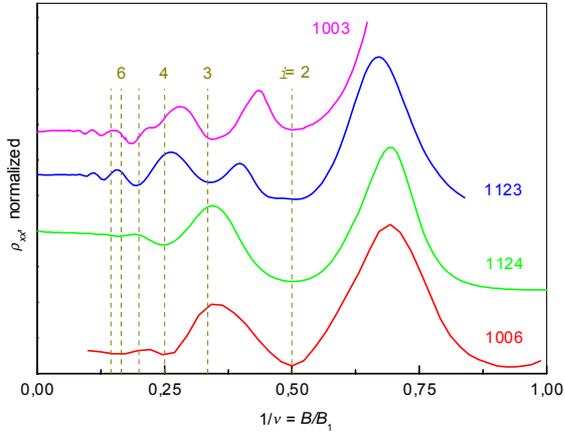

**Figure 3.** Longitudinal magnetoresistivity of the samples shown in fig.2, scaled versus inverse filling factor.

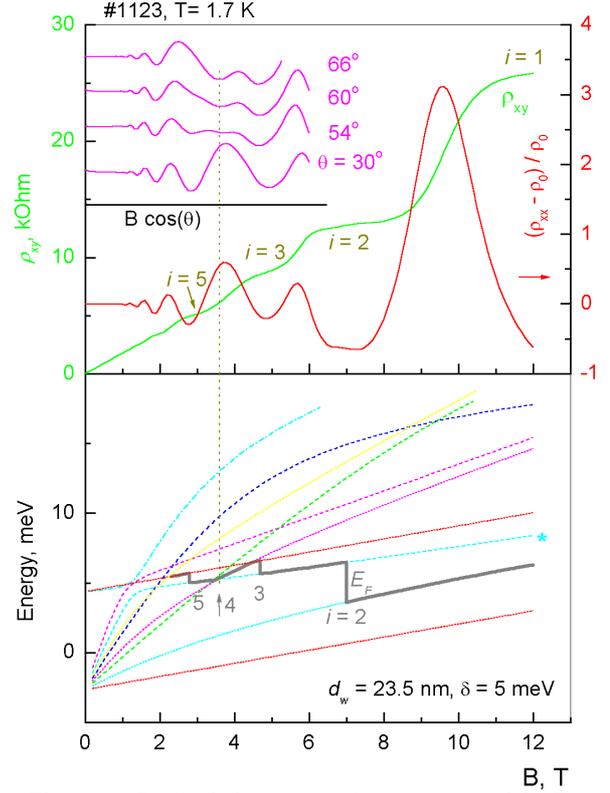

**Figure 4.** Radical changes in the structure of QHE due to involvement of the second subband.

For the samples 1006, 1124 and 1125 the plateau with $i = 4$ comes next on the low field side after the plateau with $i = 2$ and the other *even* numbered plateaus dominate. For the other two samples – 1123 and 1003 – the plateau with $i = 3$ comes next after the one with $i = 2$, the plateau with $i = 4$ is missed and the *odd* numbered plateaus dominate. These features are still more pronounced in the $\rho_{xx}(B)$ curves (figure 3). To explain this we calculated the Ge valence band energy spectra for quantization by both a magnetic field and a confinement. The model of the infinitely deep rectangular well was used [3]. Examples of these calculations are presented in figures 1(*b*) and 4. The behavior of the Fermi level is presented for the extremely sharp LLs and constant total hole densities.

As follows from the calculations, in sample 1006 the Fermi level moves within the ground confinement subband. Its levels are grouped in pairs, similar to the LL picture for a conduction band with a small spin splitting. That is why the even numbered peculiarities dominate.

In samples 1124 and 1125 the Fermi level approaches the second subband and enters its lowest LL within small intervals of field (the level marked with [*] in figures 1(*b*) and 4), but this event does not lead to essential changes in the experimental recordings.

The degree of the second subband involvement increases with $p_s d_w^2$, and the samples in the table are sorted in rows according to this parameter. Passing from sample to sample in rows of the table one can see that $p_s d_w^2$ increases abruptly for sample 1123, resulting in the Fermi level moving in it inside the second subband within a considerable part of fields (figure 4). As a result, an additional level – the lowest LL of the second subband – is imbedded in between the ground subband LLs number two and number three and divides a single step for $i = 2$ in the Fermi level motion between these two levels into two steps with $i = 2$ and $i = 3$. This is how the peculiarity $i = 3$ arises in the experimental data.

Another consequence of the imbedding the additional level is that step number 4 no longer corresponds to a transition at the distance of about the cyclotron energy between the orbitally split spin couples of the ground subband, but to a transition either onto this additional level or, if the imbedded level is lower than the Fermi level, to a transition within the spin split couples. In the later case the peculiarity numbers (the integer filling factors) are merely shifted one unit higher due to an additional level below the Fermi level. For both cases the energy distance may become much smaller than the cyclotron energy, leading particularly to a vanishing peculiarity with $i = 4$, as for sample 1123.

So the crucial role of the second confinement subband position relative to the Fermi level is evident. On the other hand, this position is very sensitive to the width and the shape of a QW that must be reflected in a high sensitivity of the structure of experimental curves to diverse changes in the system. Indeed, in sample 1003 with parameters not very different from those of sample 1123 a small peculiarity with $i = 4$ was detected. In the insert in figure 4 is shown the evolution of magnetoresistivity of sample 1123 with a tilting of magnetic field from the normal to the sample plane. The tilting introduces some changes into the LL picture that revive the peculiarity with $i = 4$, initially absent.

### 3.2. Activated magnetotransport in the quantum Hall mobility gaps

**3.2.1. The mobility gaps.** The appearance of quantized plateaux in the $\rho_{xy}(B)$ dependences with vanishing values of $\rho_{xx}$ is now commonly accepted to be caused by the existence of disorder-induced mobility gaps in the DOS of a 2D system. When the Fermi level is settled down in the gap the thermally activated behavior of $\rho_{xx}$ (or $\sigma_{xx}$) is observed due to the excitation of electrons to the narrow band of extended states (with a width $\gamma$) near the middle of a disorder-broadened LL. Direct determination of the density of localized states in the mobility gaps is possible from the measurements of Fermi energy as a function of the LL filling factor $\nu$ [5-8]. The filling factor can be changed either by the change of carrier density [5] or by the magnetic field [6-8] and the position of the Fermi level may be deduced from an analysis of the thermally activated conductivity in the magnetic field range of the Hall plateaus.

As a rule, an assumption is used that the delocalized states have discrete energies $E = E_N$ separated by the (mobility) gap $\Delta \gg k_B T$, that leads to an expression [5-8]:

$$\sigma_{xx} \sim \exp\left(-\frac{E_a}{k_B T}\right) \quad (1)$$

with $E_a = |E_F - E_N|$. In the valence band of 2D Ge, a nonlinear magnetic field dependence of LLs results in the strong inequality $\Delta \gg k_B T$ not been valid even at fields as high as $B \cong 10$ T and a more general expression should be used for the temperature dependent conductivity $\sigma_{xx}(T)$:

$$\sigma_{xx}(T) = \int \sigma(E) \frac{\partial f(E)}{\partial E} dE . \quad (2)$$

Here $f(E)$ is the Fermi-Dirac distribution function and $\sigma(E)$ is a partial conductivity at the energy $E$. For an extremely narrow band of delocalized states ($\gamma \ll k_B T$) (2) yields:

$$\sigma_{xx}(T) = \sigma_0 \frac{\gamma}{k_B T} \left.\frac{\partial f(E)}{\partial E}\right|_{E=E_a}, \quad (3)$$

where $\sigma_0$ is of the order of the minimal metallic conductivity.

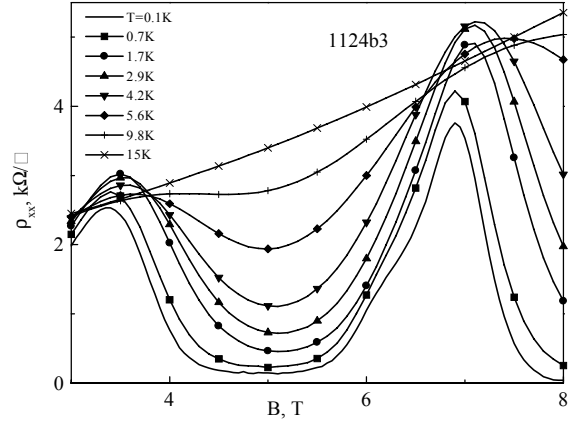

**Figure 5.** Variations of the magnetoresistivity with temperature.

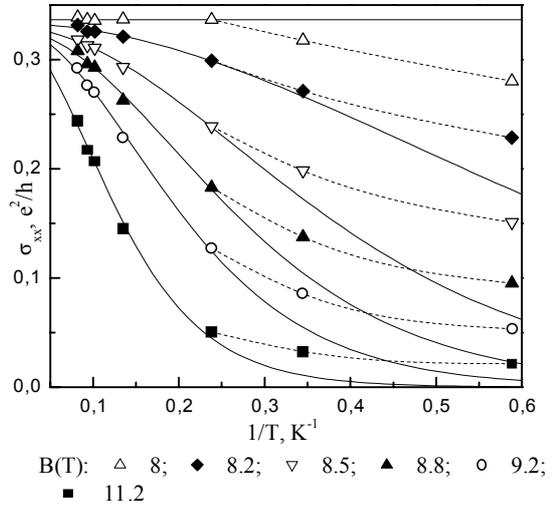

B(T): △ 8;  ◆ 8.2;  ▽ 8.5;  ▲ 8.8;  ○ 9.2;  ■ 11.2

**Figure 6.** Activation behavior of the conductivity.

In the structures investigated the magnetoresistivity was measured (fig.5) and thermally activated conductivity found for fixed values of $B$ in the quantum Hall plateau regions in the range of $T = (3 \div 15)$ K – see fig. 6. The solid curves correspond to the expression (3) with $E_a$ and $\gamma$ as a fitting parameters ($\sigma_o = 0.5 e^2/h$). Deviations of experimental points from the calculated curves for $T < 3$K (connected in fig.5 by dashed lines as a guide to the eye) are explained by variable range hopping among localized states at $E_F$, which usually dominates for sufficiently low $T$.

The activation energy $E_a$ as a function of $B$ or the filling factor $\nu$ in the vicinity of $\nu = 1$, 2 and 4 for two of the investigated samples is presented in figures 1(c) and 7. The activation energy achieves its maximum value $E_a^{max}$ at integer values of $\nu$. The mobility gap width estimated as $\Delta = 2 E_a^{max}$ is closely related to the energy separation between the adjacent LLs: $\Delta \cong |E_N - E_{N'}|$ within the uncertainty of the order of $\gamma$. In a simple parabolic band the activation energy for the integer filling factor corresponds to half the cyclotron energy:

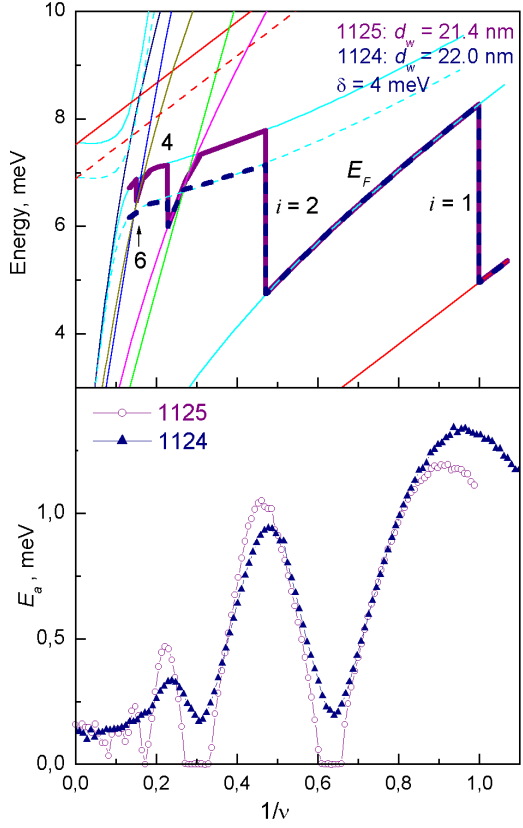

**Figure 7.** Activation energies for the sample 1125 in comparison with the data for the sample 1124 (bottom) and explanation of the difference by involvement of the second confinement subband (top: solid lines for 1125 and the dashed ones for 1124).

$E_a^{max} \cong \hbar\omega_c/2$ [7,8]. Due to a pronounced sublinearity in the $B$ dependence of the *valence band* energy levels (figures 1(*b*), 4) in complex with an interference with the second subband levels, the estimations of the cyclotron energy from the weak field Shubnikov–de Haas oscillations may strongly contradict the inter-LL distances obtained from the activation analysis in strong fields. So, $\hbar\omega_c$ should be $\cong 10$ meV at $B = 10$ T for samples 1124 and 1125 for the value $m = 0.1 m_0$ obtained from oscillations, while we have found $\Delta = (2.4 \div 2.6)$ meV and $\Delta = (1.8 \div 2.2)$ meV for $\nu = 1$ and 2, respectively, from activation conductivity.

Considering that the mobility gap corresponds to most of inter-LL distance, leaving just an infinitesimal part for the stripe of delocalized states in the middle of LLs, the values of mobility gaps so obtained could be compared with the calculated inter-LL distances. In comparison with a simple analysis of the experimental magnetoresistivity traces described in the preceding paragraph, the activation analysis yields an accurate quantitative tool to probe inter-LL distances.

An example of such an analysis for sample 1125a$_7$ is presented in figure 1, where the steps in the Fermi level motion with magnetic field due to jumps between the calculated LLs are juxtaposed with both the experimental recordings and deduced activation energies. While a pronounced step in $E_F(B)$ indicates only an existence of QH peculiarities in the $\rho_{xx}(B)$ and $\rho_{xy}(B)$, quantitative analysis is possible on the basis of deduced activation energies. Therefore, one can notice that if the whole process were developed in the first subband then the $i = 2$ mobility gap should be about 30% wider than that for $i = 1$ (figure 1(*b*)). The $i = 2$ to 1 gaps ratio may be reduced to fit the obtained activation energies if to consider the involvement of the second confinement subband. We achieved the best coincidence for the sample 1125 taking the Ge layer width $d_w = 21.4$ nm, a little higher than the nominal value of 20.0 nm.

Involvement of the second subband offers a possibility to explain the difference between activation energies in the samples 1125 and 1124 with nominally similar parameters. For the latter sample the ratio of mobility gaps for $i = 2$ to 1 is about a quarter lower than for the former (figure 7, lower part). As seen from the quantized structure of the 2D Ge valence band in the upper part of figure 7, the decrease of the mobility gap ratio in sample 1124 may be accounted for by some lowering of the second subband (dashed lines for LLs and Fermi energy in figure 7) due to an increase of the Ge layer width. The necessary correction to the Ge layer width is also small – from 21.4 to 22.0 nm – due to a strong (approximately quadratic) sensitivity of the second subband energy to the layer width.

**3.2.2. The background DOS.** From the $E_a(B)$ dependences the density of localized states in the mobility gap may be constructed [7,8]:

$$g(E_F) = \left[\frac{dE_a(n)}{dn}\right]^{-1} = \frac{\nu e}{2\pi\hbar c}\left(\frac{dE_a(B)}{dB}\right)^{-1} \quad (4)$$

In Fig. 8 are shown the typical results for the mobility gap DOS as a function of energy. Even in the middle of a gap the density of localized states was found to have an unexpectedly high value, comparable to that for a 2DHG without magnetic field $g_0 = m/(2\pi\hbar^2)$. Moreover, $g(E)$ practically does not depend on $E$ in the overwhelming part of the energy interval between adjacent LLs: $g(E) = g_c = (5 \div 8) \cdot 10^{14}$ m$^{-2}$meV$^{-1}$ for $\nu = 1$ and 2. This result is consistent with those obtained for AlGaAs/GaAs [7], InGaAs/InP [6] heterostructures and Si-MOSFET structures [5] with n-type conductivity.

In the short-range potential case theory predicts a Gaussian-like DOS

$$g_{min} \cong g_0 \exp(-\frac{\Delta^2}{4\Gamma^2}), \quad (5)$$

where $\Gamma \cong \hbar\omega_c(\mu B)^{-1/2}$. On the basis of experimental data for mobility we have found $g_{min} \sim 10^{11}$ m$^{-2}$ meV$^{-1}$, that is three order lower than the activation energy data yield for the background DOS. Thus the high background DOS between LLs cannot be explained in terms of the short-range potential.

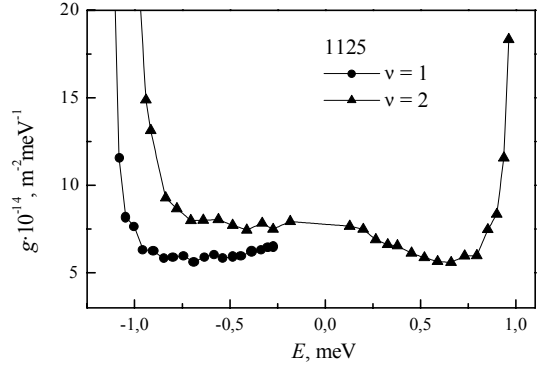

**Figure 8.** The background density of states for the sample 1125, as deduced from the activation energy. $E = 0$ corresponds to the middle of an energy interval between two LLs.

As was shown by Efros [9], in a selectively doped heterostructure the electron moves in the smooth random potential $V(x)$ created by remote impurities. This potential is caused by randomly distributed ions beyond the spacer layer. Suppose that the character scale of smooth potential variations $\bar{x} \gg l_B = (\hbar/eB)^{1/2}$. Then in a quasiclassical limit the electron energy in quantizing magnetic fields may be presented as $E_N(x_0) = \hbar\omega_c(N + ½) + V(x_0)$, with $x_0$ – the oscillator center's coordinate. Thus the smooth potential removes the degeneracy on $x_0$ and makes the LL energy dependent on spatial coordinates. Since the magnetic length $l_B \sim B^{-1/2}$, a smooth scale character of potential is manifest only in high enough magnetic fields, while under $B \rightarrow 0$ the condition must be replaced by an inequality $k_F \bar{x} \gg 1$.

If a mean value of the random potential $\bar{V} \geq \hbar\omega_c/2$ then, obviously, the states exist in the whole range between the LLs. As far as $\bar{x} \gg l_B$, $\bar{V} > \hbar\omega_c/2$ are fulfilled, the DOS is practically constant. In order to estimate the values of $\bar{x}$ and $\bar{V}$ we used the model of Shklovskii and Efros [10]. For the experimentally obtained value $g_c \cong 10^{14}$ m$^{-2}$meV$^{-1}$ the results would be: $\bar{x} \approx 1000$ Å and $\bar{V} \approx 5$ meV that yield evidence of a good implementation of the large scale potential criterion $\bar{x} \gg l_B$ ($l_B = 80$ Å at $B = 10$ T) and $\bar{V} > \Delta/2$.

## 4. Transition from single-quantum-well (SQW) to double-quantum-well (DQW) magnetotransport

A system of two closely spaced 2D layers – a DQW – creates an intriguing new physical situation since the distance between the layers can be made comparable to or less than an average distance between free carriers within a single layer. Under these conditions the interlayer correlation effects may be stronger than the intralayer ones, that results in a variety of unusual phenomena such as opening a symmetric-antisymmetric gap $\Delta_{SAS}$ dependent on magnetic field and extremely sensitive to temperature and in-plain magnetic field [11], stabilization of some phases, forbidden or hardly realizable in a single 2DEG, such as the even-denominator fractional QHE [12], the Wigner crystal insulating state [13] etc.

A DQW system may be created in a sufficiently wide and intensively populated single well [14]. The idea is that, when electrons are introduced in a wide QW, the electrostatic repulsion between the electrons forces them into a stable configuration in which two 2DEG's are formed at the well sidewalls. A major advantage of this system over a conventional DQW with a barrier deliberately introduced inside a QW is the minimization of alloy scattering since the barrier between the two 2DEG's is of the same material, without heterointerfaces. Also, varying the carrier density in the well can change both $\Delta_{SAS}$ and inter-sublayer distance.

So far almost all of the research in the DQWs has been performed in the electron systems. The hole system offers some new properties in the DQW: particularly the large value of the heavy hole mass allows us to achieve easily the configuration with Coulomb coupled hole gases without tunneling between them.

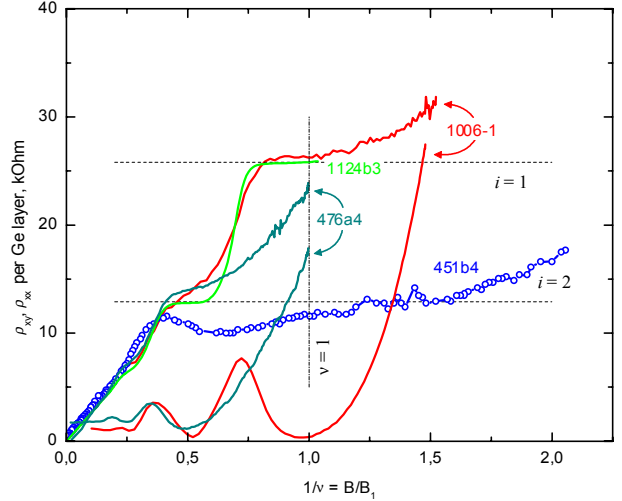

**Figure 9.** Transformations of the QHE with transition from SQW to DQW magnetotransport.

To try the phenomenon of the hole gas segregation in two sublayers in our heterosystem, we have investigated a group of several samples with wider QWs represented in the table by samples 451 and 476. The experimental results for these samples are shown in fig.9 in comparison with results for the samples discussed in section 3.

As is seen in this figure, the results for samples 451 and 476 differ considerably from those with a narrow Ge layer (see also figures 2 and 3) in that the QH state with $i = 1$ is absent here while it distinctly exists for the others. Considering that the spin splitting in the valence band of 2D Ge is as high as nearly a half of the orbital one [3] and that the $i = 1$ QH state is well resolved in the narrow well samples, the missing of the $i = 1$ QH state in the wide wells means unambiguously that conductivity in each Ge layer is via two parallel 2D sublayers here.

This conclusion is confirmed by our calculations of the potential profile, energy levels and wave functions (made self consistently from a system of Schrodinger and Poisson equations) – see fig.10. For sample 1006 all the levels are higher than the bottom bending amplitude, and the ground wave function is uniformly distributed within the well, in contrast to the results for samples 451 and 475/476. In samples 475/476 the two lowest levels $E_0$ and $E_1$ coincide and their wave functions are confined within triangular wells next to the well walls, indicating the complete separation of the hole gas in two sublayers.

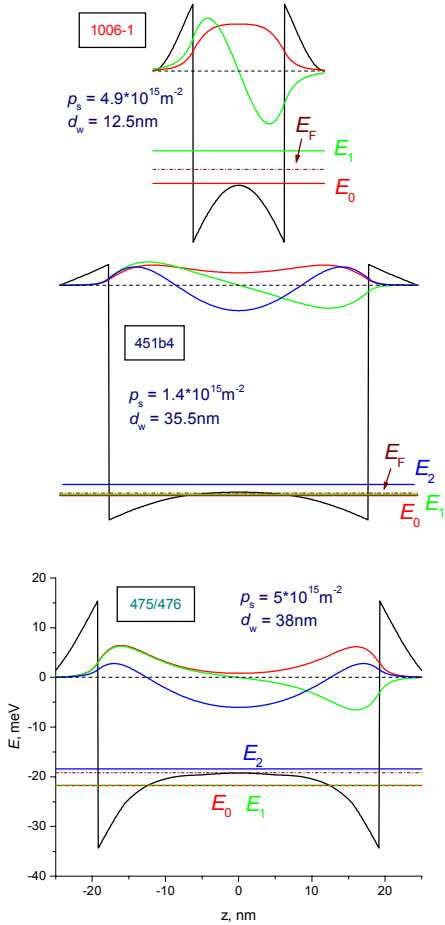

**Figure 10.** Calculated potential profiles, energy levels and wave functions of the samples researched.

A positive longitudinal magnetoresistance has been observed in the weakest fields only for the 475/476 group of samples (figure 11), indicating participation of two kinds of carriers with different mobilities $\mu_i$. To determine the characteristics of these two components of conductivity, we have fitted our data for positive magnetoresistivity and concomitant slightly sublinear Hall magnetoresistance to simple Drude-Lorenz formula [15]:

$$\rho_{xx} = (D_1 + D_2)/[(D_1 + D_2)^2 + (A_1 + A_2)^2]$$
$$\rho_{xy} = -(A_1 + A_2)/[(D_1 + D_2)^2 + (A_1 + A_2)^2]$$

where $D_i = p_i e\mu_i/(1 + \mu_i B)$ is the diagonal term of the conductivity matrix for layer $i$, $A_i = \mu_i B D_i$ is the off-diagonal term and $p_i$ is the sheet hole density.

The results (figure 11) indicate that either with equal hole densities $p_1 = p_2$, or (with slightly better agreement) with $p_1 \neq p_2$, the mobilities of the two components are about a factor of two different. The difference is probably due to the different quality of the normal and inverted interfaces that border the QW. So, up to 40-fold difference in mobility was found for electron sublayers located at the normal and inverted interfaces of the $GaAs/Ga_{1-x}Al_xAs$ QW [16]. Bad mobility near the inverted interface was attributed mainly to the scattering on Si dopants floating up from the lower $Ga_{1-x}Al_xAs$ barrier during the growth.

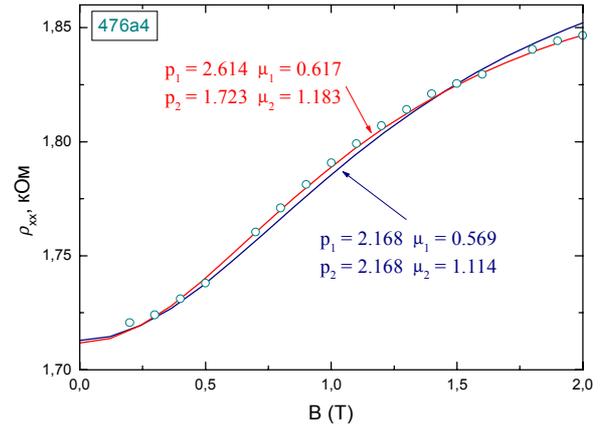

**Figure 11.** Positive magnetoresistance (circles) and fitted curves.

For such a pronounced difference in mobilities, how could the QH plateau be observed close to the fundamental value $\rho_{xy} = h/ie^2$, $i = 2$, (sample 476 in fig.9) for the two parallel sublayers? To find the answer, the equivalent circuit of the experiment was analyzed – see figure 12. Due to curved equipotential lines in the vicinity of the current leads (figure 12(*a*)) the Hall resistivity of each sublayer $R_{xyi}$ ($= h/e^2$ in the field range of the first plateau) is switched in series into the sublayer circuit (figure 12(*b*)) [17]. Then, the potential contacts to the sample, which penetrate the whole depth of a volume with conducting multilayers, would connect the Hall resistances of the sublayers in parallel (figure 12(*c*)). Therefore the resultant Hall resistivity of the Ge layer will depend on the values of the sublayer Hall resistances only, no matter how different the mobilities (i.e. $R_{xxi}$) are. That would yield a plateau at $R_{xy} = h/2e^2$ for not very different hole densities in the sublayers, as has been measured for the Ge layer.

The most unusual results were obtained on a group of three samples within the intermediate range of sublayer separations, with $d_w = 34 \div 41$ nm, comparable to the previously described sample group, but with the lower hole densities $p_s = (1 \div 2) \cdot 10^{15}$ m$^{-2}$ (see sample 451b4 in figure 9). Each of these three samples exhibited an unusually wide plateau in $\rho_{xy}(B)$ close to the value 13 kOhm, which corresponds to $i = 2$. We interpret the latter as a manifestation of a double-layer quantized Hall insulator state stabilized by the interlayer correlations [18].

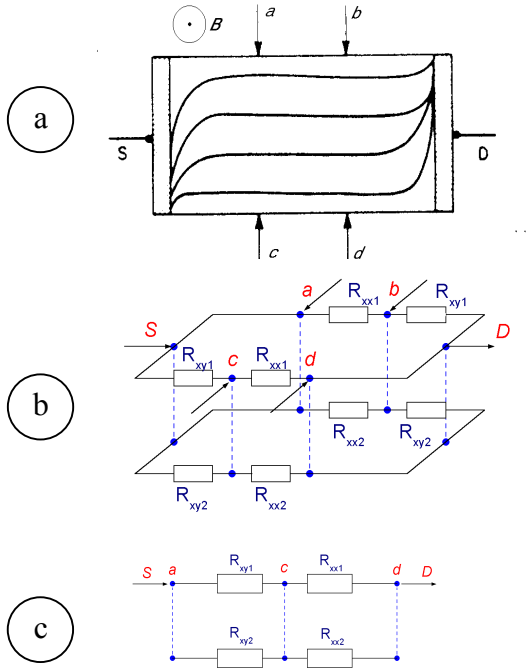

**Figure 12.** Equivalent circuit of the double layer system in the quantum Hall regime.
a) curved potentials in the 2D layer;
b) distribution of the Hall- and longitudinal resistances in the layers;
c) parallel connection of the layer resistances

In the calculated results for the sample 451 (figure 10) the first two levels $E_0$ and $E_1$, as well as the Fermi level, are lower but very close to the bottom bending amplitude. Even for the level $E_0$ the ground state wave function $\Psi_0$ is far from zero in the center of the well. The differences from the samples 475/476 are due to much lower hole density in the well. All these features indicate the possibility of existence of two coupled hole sublayers in the Ge layer. Considering the large hole mass, even a lowest barrier suppresses tunneling, that is reflected in the small splitting between $E_0$ and $E_1$ levels. However, a small depth of a $\Psi_0$ minimum in the center of the well indicates that the effective distance between the sublayers might be small enough, for the interlayer correlations were substantial.

## 5. Concluding remarks

Reach information concerning QWs in the p-Ge$_{1-x}$Si$_x$/Ge heterosystem may be obtained from the study of magnetotransport in the quantum Hall regime. Under conditions when the second subband is involved in magnetotransport, crossings of LLs originating from different subbands result in the existence or absence of certain peculiarities in experimental recordings. When the second subband just starts to be populated, the changes in distances between populated levels may be deduced from the activated behavior of the magnetoresistance at magnetic fields within the quantum Hall plateaux. Both of these phenomena are extremely sensitive to the well potential profile.

Activated behavior in the quantum Hall regime also yields the background DOS that characterizes random potential distribution in the system.

The quality of the normal and inverted interfaces may be quantitatively compared in sufficiently wide and (or) populated wells, when the hole gas in the well separates into two sublayers located at the sidewalls. In the structures investigated mobilities different by a factor of two were deduced.

It was shown that unique features of DQWs might be realized in the p-Ge$_{1-x}$Si$_x$/Ge heterosystem in certain ranges of Ge layer widths and hole densities. Particularly, the quantized Hall insulator state has been found to exist just at the transition from the integrated 2DHG in a QW to a system of separated 2D sublayers in a Ge layer.

## Acknowledgments

This work was supported by Russian Foundation for the Basic Researches, projects 99-02-16256 and 98-02-17306.